\documentclass[preprint,showpacs,preprintnumbers,amsmath,amssymb]{revtex4}

\usepackage{graphicx}% Include figure files
\usepackage{dcolumn}% Align table columns on decimal point
\usepackage{bm}% bold math
\usepackage{cases}
\usepackage[usenames]{color}

\begin{document}

\preprint{}

\title{Phase Separation in Bose-Bose Mixtures in an Optical Lattice}% Force line breaks with \\

\author{Takeshi Ozaki}
\email{j1209702@ed.kagu.tus.ac.jp}
% \altaffiliation{Department Physics, Faculty of Science, Tokyo University of Science, 1-3 Kagurazaka, Shinjuku-ku, Tokyo, 162-8601, Japan}%Lines break automatically or can be forced with \\
\author{Tetsuro Nikuni}%
 \email{nikuni@rs.kagu.tus.ac.jp}
\affiliation{%
Department Physics, Faculty of Science, Tokyo University of Science, 1-3 Kagurazaka, Shinjuku-ku, Tokyo, 162-8601, Japan}%

%\affiliation{

\date{\today}% It is always \today, today,

\begin{abstract}
We study the ground-state properties of mixtures of strongly interacting bosonic atoms in an optical lattice.
Applying a mean-field approximation to the Hubbard model for Bose-Bose mixtures, we calculate the densities and superfluid order parameters for both species.
Due to the repulsive interaction between the two species, the system exhibits phase separation.
First, in the extreme limit of the zero-hopping case, we derive analytical expressions for the phase boundaries.
In particular, we derive the conditions for phase separation in the Mott insulator phase.
We find that the conditions for the phase separation depend on the on-site interactions as well as the occupation numbers.
In particular, we show that the coexisting state appears by varying the on-site inter-species interaction.
We also show the phase diagram of the finite hopping case.
Second, we calculate the spatial density profile of $^{87}$Rb-$^{41}$K mixtures in the combined potential of a parabolic trap and an optical lattice using the local density approximation.
We fixed the number of $^{87}$Rb and varied the number of $^{41}$K, and used the parameters estimated by experiments.
We show that the phase separated $^{87}$Rb-$^{41}$K mixtures distribute like in a parabolic trap case.
Furthermore, we find that phase separated mixtures distribute a nesting structure.
\end{abstract}

\maketitle
\section{INTRODUCTION}
Systems of ultracold atoms in an optical lattice have attracted attention for studies of strongly correlated quantum matter.
One of the most interesting phenomena exhibited by ultracold Bose atoms in an optical lattice is the superfluid (SF) to Mott insulator (MI) phase transition, which has been experimentally observed \cite{nature2002}. 
The properties of this system are well-captured by the Bose-Hubbard model \cite{jakschPRL81,OostenPRA2001,LuPRA2006}. 
In a shallow optical lattice, bosonic atoms are in an SF phase,
wherein atoms have long-range phase coherence and number fluctuations.
On the other hand, atoms are in an MI phase in a deep optical lattice,
wherein the filling factor is fixed to an integer and phase coherence disappears.

By the addition of a second atomic species, the system exhibits rich quantum phases.
Some experimental groups are working on Fermi-Fermi~\cite{Nature_455_204,PRL_102_020405}, Bose-Fermi~\cite{PRL_96_180402,PRL_89_150403,PRL_102_030408}, and Bose-Bose\cite{Catani,Science_319_295} mixtures in optical lattices.
Recently, the Florence group experimentally showed $^{87}$Rb-$^{41}$K mixtures trapped in an optical lattice \cite{Catani}.
The observed interference pattern revealed that the visibility of $^{87}$Rb is lower than the pure $^{87}$Rb case.

The properties of Bose-Bose mixtures in an optical lattice have been studied theoretically using various approaches, such as by applying the Hubbard model using the perturbation theory \cite{HanPL332}, the Gutzwiller approach \cite{PRL_100_240402}, the time evolving block decimation method \cite{PRA_80_023619}, and quantum Monte Carlo simulations \cite{RoscildePRL2007}.
In these previous studies, a supersolid phase and a counterflow superfluid phase have been predicted to exist, assuming that the two components have the same mass.
On the other hand, it is important to consider the difference in components to investigate experimentally obtained two-species mixtures in an optical lattice.

In this study, we have investigated two-species Bose mixtures in an optical lattice.
In particular, we focus on the conditions for phase-separation.
We have considered realistic parameter sets for the $^{87}$Rb-$^{41}$K mixture. 
Specifically, we consider mixtures of different bosonic atoms with different masses, on-site interactions, and hopping amplitudes. 
Our main purpose is to clarify superfluid order parameters in a combined potential of a parabolic potential and an optical lattice.
We apply the mean-field approximation and include the effects of trap potential by the local density approximation.
We calculate spatial profiles of the densities and the superfluid order parameters.

This paper is organized as follows: 
In section~\ref{Bose Hubbard Hamiltonian}, we explain the mean-field approximation and the numerical method used.
In section~\ref{PHASE-SEPARATION IN MI PHASE}, we present the conditions for the spatial phase separation of Bose-Bose mixtures in an optical lattice.
In order to gain qualitative insight, we also present the phase diagram in a zero-hopping limit. 
In section~\ref{sec:results}, we present the phase diagram and the density profiles of $^{87}$Rb-$^{41}$K mixtures in a combined potential of a parabolic trap and an optical lattice. 
We will show that the phase separated $^{87}$Rb-$^{41}$K mixtures distribute a nesting structure.

\section{Mean-field approximation for the Bose-Hubbard Hamiltonian}
\label{Bose Hubbard Hamiltonian}
In this section, we describe the mean-field approximation for the Bose-Bose mixture in the absence of a parabolic potential.
We consider two-component bosonic-atom mixtures trapped in an optical lattice at zero temperature.
We use the following Bose-Hubbard Hamiltonian for the Bose-Bose mixtures;
\begin{align}
\label{eqn:bbmhm}
H  =&\sum_{\alpha=1,2}\left[-t_{\alpha} \sum_{\langle i,j\rangle} (\hat{b}_{\alpha i} ^\dag \hat{b}_{\alpha j} + \hat{b}_{\alpha j}^\dag\hat{b}_{\alpha i}) + \frac{U_{\alpha}}{2}\sum_{i} \hat{n}_{\alpha i}(\hat{n}_{\alpha i}-1) -\mu_{\alpha}\sum_{i} \hat{n}_{\alpha i}\right]  \nonumber \\
& +U_{12} \sum _i \hat{n}_{1i}\hat{n}_{2i},
\end{align}
where $\hat{b}_{\alpha i} ^\dag (\hat{b}_{\alpha i})$ is the creation (annihilation) operator of the $\alpha(=1$ or $2)$ component at site $i$, $\hat{n}_{\alpha i}\equiv \hat{b}_{\alpha i} ^\dag \hat{b}_{\alpha i} $ is the number operator, and $\langle i,j \rangle$ denotes the sum over the nearest neighbor sites.
$t_{\alpha}$, $U_{\alpha}$ and $\mu_{\alpha}$ are the hopping amplitude, the on-site intra-species interaction and the chemical potential respectively, and $U_{12}$ is the on-site inter-species interaction.
Throughout this paper, we consider the repulsive on-site inter-species interaction, i.e. $U_{12}>0$.

Using a mean-field approximation \cite{OostenPRA2001}, we decouple the hopping term as
\begin{align}
\label{eqn:MF1}
\hat{b}_{\alpha i} ^{\dag} \hat{b}_{\alpha j} \simeq & \langle \hat{b}_{\alpha i} ^{\dag } \rangle \hat{b}_{\alpha j} +\hat{b}_{\alpha i} ^{\dag} \langle \hat{b}_{\alpha j} \rangle - \langle \hat{b}_{\alpha i} ^{\dag }\rangle \langle \hat{b}_{\alpha j} \rangle \nonumber \\
=&\ \phi_{\alpha} \hat{b}_{\alpha j} + \hat{b}_{\alpha i}\phi_{\alpha}  - \phi_{\alpha}^2 ,
\end{align}
where $\phi_{\alpha} =\langle \hat{b}_{\alpha i}^\dag\rangle = \langle \hat{b}_{\alpha i} \rangle$ is the superfluid order parameter of the $\alpha$ component.
Within this approximation, the mean-field Hamiltonian can be written as a sum over single-site terms, $H\simeq H_\text{MF}=\sum _i H_i$, where the single-site Hamiltonian is given by
\begin{align}
\label{hamiltonian:i-site}
H_i =& \sum_{\alpha}\left[-zt_{\alpha}\phi_{\alpha} (\hat{b}_{\alpha i} ^\dag + \hat{b}_{\alpha i})+\frac{U_{\alpha}}{2} \hat{n}_{\alpha i} (\hat{n}_{\alpha i} -1) - \mu_{\alpha} \hat{n}_{\alpha i} +zt_{\alpha} \phi_{\alpha}^2\right] \nonumber \\
& +U_{12}\hat{n}_{1i}\hat{n}_{2i},
\end{align}
where $z$ is the coordination number.

We can calculate the densities and the superfluid order parameters using the mean-field Hamiltonian (\ref{hamiltonian:i-site}).
We employ the same basic method used by Lu and Xaing \cite{LuPRA2006}, adapted to the case of Bose-Bose mixtures.
The ground energy $E_0$ and ground state wavefunction $|\psi_0\rangle$ of Bose-Bose mixtures can be obtained by diagonalizing the Hamiltonian $H_i$ in the occupation number basis $\{ |n_1,n_2 \rangle \}$ truncated at finite values $n_{t1}$ and $n_{t2}$.
In the present work, we set $n_{t1}=n_{t2}=n_{t}$ and use a sufficiently large $n_t$ so that the results do not depend on $n_t$.
For a given $U_{\alpha}$, $U_{12}$, $t_{\alpha}$, and $\mu_{\alpha}$, the superfluid order parameters can be determined by minimizing $E_0$.
The region with nonzero $\phi_{\alpha}$ is identified as the superfluid phase while the region with $\phi_{\alpha}=0$ is identified as the Mott-insulator phase.
After determining $\phi_{\alpha}$, we obtained the superfluid densities $\rho_{s\alpha}$ and average densities $\rho_{\alpha}$ as
\begin{align}
\rho_{s\alpha}=\phi_{\alpha}^2,\\
\rho_{\alpha}=\langle \hat{n}_{\alpha} \rangle.
\end{align}

\section{PHASE-SEPARATION IN MI PHASE}
\label{PHASE-SEPARATION IN MI PHASE}
Before presenting detailed results of the mean-field theory, we clarify the condition for the phase separation.
Here, we consider a zero-hopping limit $t_1(t_2)\rightarrow0$ in order to obtain a qualitative understanding of the phase separation.
We denote the energy of $|n_1,n_2\rangle$ per site by $E_{n_1,n_2}$, and we refer to this MI state as the $(n_1,n_2)$ state.
An explicit expression for this energy is given by
\begin{align}
\label{eqn:mu}
E_{n_1,n_2}=\frac{U_1}{2}n_1(n_1-1)+\frac{U_2}{2}n_1(n_2-1)+U_{12}n_1n_2-\mu_1n_1-\mu_2n_2.
\end{align}
For the $(n_1,n_2)$ state to be the ground state, one must satisfy $E_{n_1,n_2}<E_{n_1\pm1,n_2},E_{n_1,n_2\pm1}$.
This leads to the following relations between the chemical potentials and the number of $\alpha$ components;
\begin{equation}
\label{eqn:che_num}
\begin{split}
U_1(n_1-1)+U_{12}n_2 < \mu _1 < U_1n_1+U_{12}n_2,  \\
U_2(n_2-1)+U_{12}n_1 < \mu _2 < U_2n_2+U_{12}n_1.
\end{split}
\end{equation}
If $\mu_{\alpha}<0$, there are no particles of the $\alpha$ component.

We now consider the dependence on $U_{12}$.
In the limit $U_{12}\to0$, the occupation numbers of two components should be independently determined by Eq.~(\ref{eqn:che_num}).
On the other hand, in the opposite limit $U_{12}\to\infty$, two components cannot coexist at the same site, so that the mixtures are phase-separated, i.e. the ground state is either the $(n_1,0)$ state or the $(0,n_2)$ state.
In this limit, the phase boundary between two phase-separated states is analytically given by:
\begin{eqnarray}
\label{eqn:bun}
\mu_2 = \frac{n_1}{n_2}\mu_1 + \frac{U_2}{2}(n_2-1) - \frac{U_1}{2}\frac{n_1}{n_2}(n_1-1),
\end{eqnarray}
where $n_1$ and $n_2$ must be satisfied Eq.~(\ref{eqn:che_num}).

We now derive the condition for the coexisting state.
With decreasing magnitude of $U_{12}$ from $U_{12}\to\infty$ limit, the coexisting state $(n_1,1)$ or $(1,n_2)$ first appears.
In order for the $(n_1,1)$ state to appear between the $(n_1,0)$ state and the $(0,n_2)$ state, one must satisfy $E_{n_1,1}<E_{n_1,0}$.
Therefore, from Eq.~(\ref{eqn:mu}) we obtain
\begin{align}
\label{eqn:rel_coex}
\mu_2>U_{12}n_1.
\end{align}
The intersection of $\mu_2=U_{12}n_1$ and Eq.(\ref{eqn:bun}) is given by
\begin{align}
\mu_1=U_{12}n_2+\frac{U_1}{2}(n_1-1)-\frac{U_2}{2}\frac{n_2}{n_1}(n_2-1).
\end{align}
Using Eq.~(\ref{eqn:che_num}), we find the condition for the appearance of the $(n_1,1)$ state to be:
\begin{align}
\label{eqn:PBcon2}
U_{12}<\frac{1}{n_1n_2}\left[\frac{U_1}{2}n_1(n_1+1)+\frac{U_2}{2}n_2(n_2-1)\right].
\end{align}
From a similar calculation, the condition for the appearance of the $(1,n_2)$ state between the $(n_1,0)$ state and the $(0,n_2)$ state is given by
\begin{align}
\label{eqn:PBcon3}
U_{12}<\frac{1}{n_1n_2}\left[\frac{U_1}{2}n_1(n_1-1)+\frac{U_2}{2}n_2(n_2+1)\right].
\end{align}
These results are consistent with those of previous studies \cite{PRL_100_240402,PRA_76_013604}.
If we set $U_1=U_2=U$, as in \cite{PRA_76_013604}, the condition for the coexisting state is given as $U_{12}<U$.
We note that the condition for the coexisting state in the MI phase depends on the on-site interactions as well as the occupation numbers.

\section{NUMERICAL RESULTS}
\label{sec:results}
In this section, we present numerical results of the mean-field calculation, considering the $^{87}$Rb-$^{41}$K mixture used in the experiment of Ref.~\cite{Catani}.
We assume that $\alpha=1$ denotes $^{87}$Rb and $\alpha=2$ denotes $^{41}$K.
We calculate the densities and the superfluid order parameters using the method explained in Sec.~\ref{Bose Hubbard Hamiltonian}.
We estimate the on-site interactions and hopping amplitudes using the experimental data given in \cite{Catani}:
lattice laser wavelength $\lambda_L=1064$nm, scattering length $a_{Rb}=99a_0$, $a_{K}=65a_0$, $a_{Rb-K}=169a_0$, where $a_0$ is the Bohr radius.
These results indicate that the on-site interactions are $U_1:U_2:U_{12}=1:0.34:1.85$, the hopping amplitudes are $t_1:t_2=1:9.39$, and $U_1=40zt_1$.

\subsection{No hopping limit}
\begin{figure}[hts]
\begin{tabular}{cc}
\begin{minipage}{0.5\hsize}
\begin{center}
\includegraphics[width=\hsize]{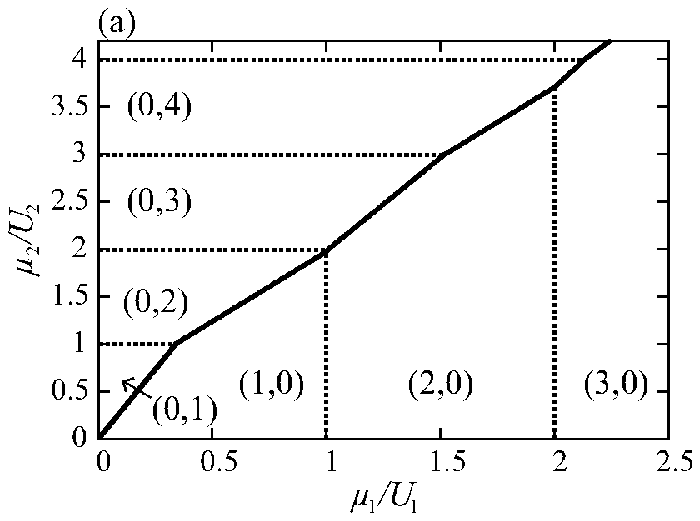}%
\end{center}
\end{minipage}
\begin{minipage}{0.5\hsize}
\begin{center}
\includegraphics[width=\hsize]{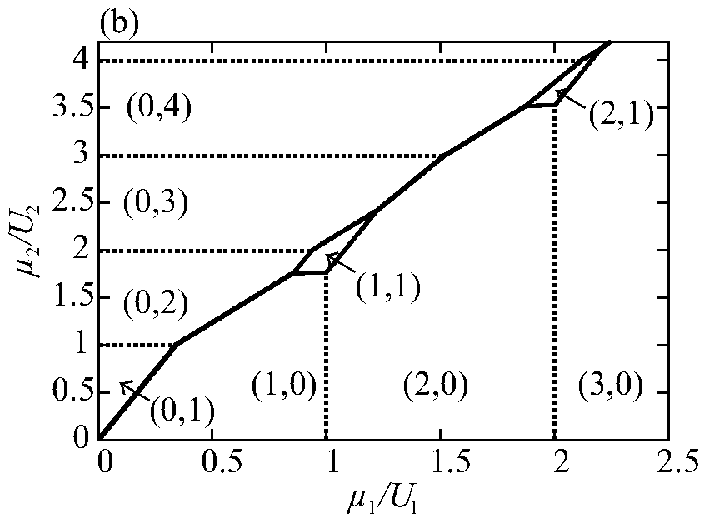}
\end{center}
\end{minipage}
\end{tabular}
\caption{\label{Fig:bunN}The phase boundary of ($n_1,n_2$) states in the zero-hopping limit  with (a) $U_1:U_2:U_{12}=1:0.34:1.85$, (b)$U_1:U_2:U_{12}=1:0.34:0.6$. The solid line shows the boundary between $^{87}$Rb and $^{41}$K. The dotted lines show the boundary between the same species with different occupation numbers.}
\end{figure}

As an illustration, we consider the ground state in the no hopping limit, setting $t_{\alpha}=0$.
In Fig.~\ref{Fig:bunN}, we plot the phase boundaries of the $\mu_1/U_1$-$\mu_2/U_2$ plane separating the regions with different occupation numbers.
Figure~\ref{Fig:bunN}(a) shows the phase diagram obtained for the parameters described above.
Since the on-site interactions do not satisfy the condition for coexistence, the mixture is always phase separated.
Note that the slope of the phase boundary is different for different occupation numbers.
The variation of the slope of the phase boundary is due to the difference of the on-site intra-species interactions $U_1$ and $U_2$.

We next consider the effects of changing the value of $U_{12}$.
Experimentally, the magnitude of $U_{12}$ can be controlled by the Feshbach resonance \cite{PRL_100_210402}.
In Fig.~\ref{Fig:bunN}(b), we plot the phase boundary for a weaker inter-species interaction $U_{12}=0.6U_1$.
In this case, the condition for coexistence (\ref{eqn:PBcon2}) is satisfied, however, the condition (\ref{eqn:PBcon3}) is not satisfied.
Contrary to Fig.~\ref{Fig:bunN}(a), the coexisting phase appears in the region enclosed by the solid line in Fig.~\ref{Fig:bunN}(b).
However, most regions are still in the separated phases $(n_1,0)$ state or the $(0,n_2)$ state.

\subsection{Finite hopping case ($t_{\alpha}\neq0$)}
\begin{figure}[hts]
\includegraphics[width=0.5\hsize]{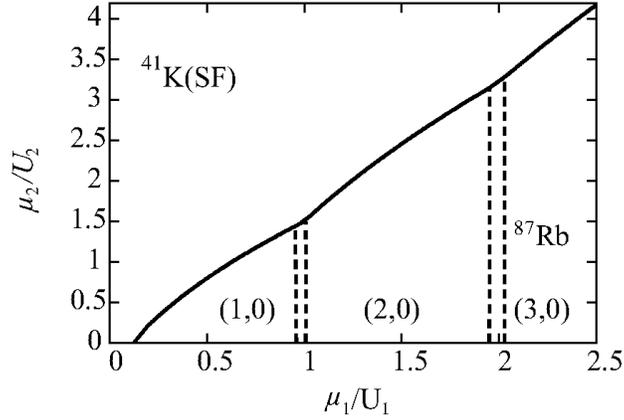}
\caption{\label{Fig:PDtneq0}Phase diagram of $^{87}$Rb-$^{41}$K mixtures in the $\mu_1/U_1$-$\mu_2/U_2$ plane. The solid line represents the phase boundary of the $(n_1,0)$ state and the $(0,n_2)$ state. The dashed lines represent the phase boundary of the SF phase and the MI phase.
}
\end{figure}

We now consider the phase diagram for finite hopping amplitudes.
In particular, we set the parameters for $^{87}$Rb to be the MI phase and $^{41}$K to be the SF phase.
In Fig.~\ref{Fig:PDtneq0}, we plot the phase diagram of the $^{87}$Rb-$^{41}$K mixture at finite hopping.
We find that the ground-state remains phase-separated, and the $^{41}$K(SF) region becomes larger while the $(n_1,0)$ region becomes smaller than the $t_{\alpha}=0$ case.
Since the SF phase appears in the ground-state, the phase boundary between the $(n_1,0)$ state and the $^{41}$K(SF) state now becomes a smooth curve in contrast to Fig.~\ref{Fig:bunN}(a).

\subsection{Effect of parabolic potential}
We now consider the effect of a parabolic potential.
We will show that an interesting feature arises in the density profiles of $^{87}$Rb-$^{41}$K mixtures in a combined potential.
We fix the number of $^{87}$Rb and control the number of $^{41}$K by changing the chemical potential $\mu_2$.
In order to include the effect of the trap potential, we apply a local density approximation (see Ref.~\cite{jakschPRL81}).
That is, we set the local chemical potential for each species as $\mu_{\alpha}^{\text{eff}}(\textbf{r}_i) = \mu_{\alpha}^0 - \epsilon_{\alpha}(\textbf{r}_i)$, where $\mu_{\alpha}^0$ is the chemical potential at the center of a parabolic potential and $\epsilon_{\alpha}(\textbf{r}_i)=\frac{zt_1}{2}\frac{m_{\alpha}}{m_1}\textbf{r}_i^2$ is the parabolic trap contribution for each species.
It should be noted that we consider the difference in the masses of the two species.

\begin{figure}[hts]
\begin{tabular}{cc}
\begin{minipage}{0.5\hsize}
\begin{center}
\includegraphics[width=\hsize]{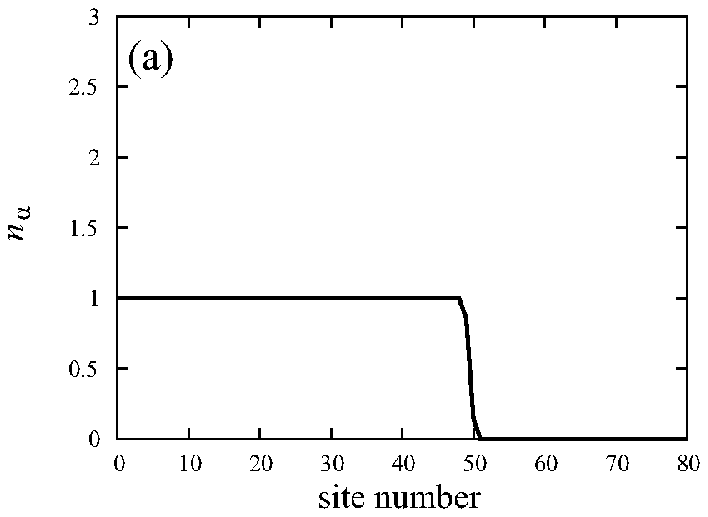}
\end{center}
\end{minipage}
\begin{minipage}{0.5\hsize}
\begin{center}
\includegraphics[width=\hsize]{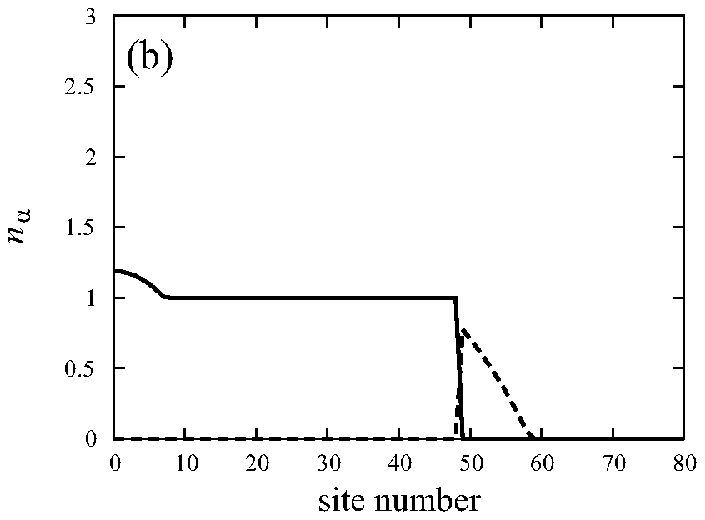}
\end{center}
\end{minipage}
\end{tabular}
\begin{tabular}{cc}
\begin{minipage}{0.5\hsize}
\begin{center}
\includegraphics[width=\hsize]{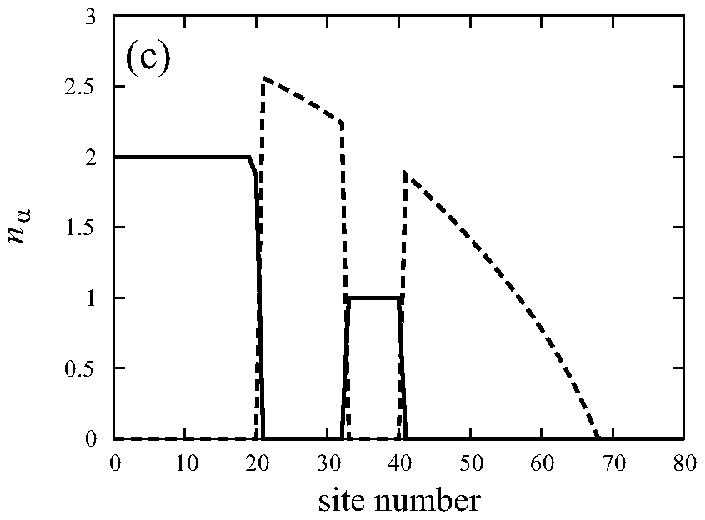}
\end{center}
\end{minipage}
\begin{minipage}{0.5\hsize}
\begin{center}
\includegraphics[width=\hsize]{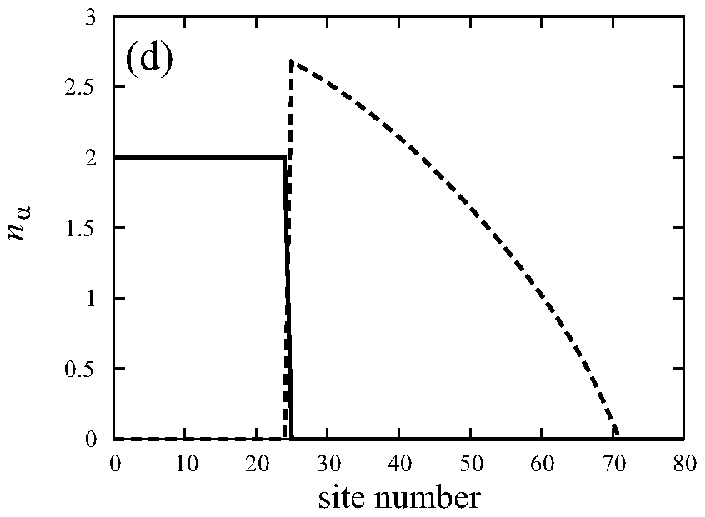}
\end{center}
\end{minipage}
\end{tabular}
\caption{\label{Fig:DP}Density distribution of $^{87}$Rb-$^{41}$K mixtures in a combined potential. (a)$\mu_2^0/zt_1=0$, (b)$\mu_2^0/zt_1=16$, (c)$\mu_2^0/zt_1=25$, and (d)$\mu_2^0/zt_1=28$. The solid line is the occupation number of $^{87}$Rb, and the dashed line is the occupation number of $^{41}$K.
}\end{figure}

The variation in the density profiles is plotted in Figs.~\ref{Fig:DP} (a)-(d).
In Fig.~\ref{Fig:DP}(a), we plot the density of a pure $^{87}$Rb gas, i.e. $^{41}$K atoms are absent.
The density of $^{87}$Rb in a parabolic potential exhibits the so-called wedding-cake profile, with a plateau corresponding to the MI ($n_1=1$) domain at the center of a parabolic potential and the SF state at the outside of the MI domain.
Figure~\ref{Fig:DP}(b) shows the density profiles of the $^{87}$Rb and $^{41}$K mixtures for $\mu_2^0/zt_1=16$.
Adding the $^{41}$K component under the conditions for the phase separation (\ref{eqn:PBcon2}) and (\ref{eqn:PBcon3}), we see that $^{87}$Rb exists in the center of a parabolic trap, which is surrounded by the $^{41}$K component.
An analogous structure is also found in the density profile of the BEC mixture without the optical lattice in a parabolic trap \cite{PethickSmith2nd}.
We note that the $^{41}$K component in the outer region pushes the $^{87}$Rb component.
Thus, the Mott domain of the $^{87}$Rb component at the center of the trap, seen in Fig.~{Fig:DP} (a), is turned into the SF region.
This result is consistent with the results obtained using the Gutzwiller approach \cite{PRL_100_240402}.
We find that the Mott core changes to the superfluid at the center of a parabolic trap and $^{41}$K exist outside the $^{87}$Rb component.

Upon further increasing the population of the $^{41}$K component, the $^{87}$Rb component of the SF state at the center of the trap is compressed, and thus the $n_1=2$ $^{87}$Rb Mott plateau appears at the center of the parabolic potential.
However, not all the$^{87}$Rb components change to the $n_1=2$ MI state and the $n_1=1$ MI state remains.
Moreover, the $^{41}$K components penetrate between the $n_1=2$ and $n_1=1$ $^{87}$Rb Mott plateaus.
Therefore, as can be seen in Fig.~\ref{Fig:DP}(c), the density profile becomes a nesting structure with alternately arranged $^{87}$Rb, and $^{41}$K from the center to the outside of a combined trap.
This nesting structure is a peculiar feature of a Bose-Bose mixture trapped in an optical lattice.
Upon further increasing the population of the $^{41}$K component, the $n_1=1$ $^{87}$Rb Mott plateau decreases and the $n_1=1$ $^{87}$Rb Mott plateau increases.
In Fig.~\ref{Fig:DP}(d), we plot the results for $\mu_2^0/zt_1=28$.
We find that the nesting structure seen in Fig.~\ref{Fig:DP}(c) disappears.
As in Fig.~\ref{Fig:DP}(b), all of the $^{41}$K exists outside the $^{87}$Rb component again.

\begin{figure}[hts]
\includegraphics[width=0.5\hsize]{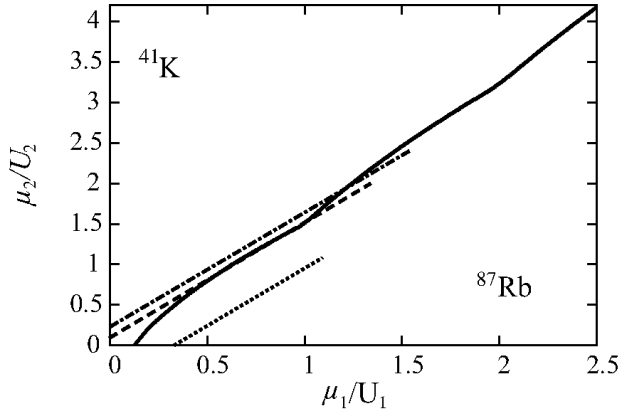}
\caption{\label{Fig:PDtneq0_2} Phase diagram of $^{87}$Rb-$^{41}$K mixtures in the $\mu_1/U_1$-$\mu_2/U_2$ plane. The solid line is the phase boundary of the $(n_1,0)$ state and the $(0,n_2)$ state.
We plot the trajectories of the chemical potential for $\mu_2^0/zt_1=16$ (dotted line), $\mu_2^0/zt_1=25$ (dashed line), and $\mu_2^0/zt_1=28$ (dash-dotted line) respectively.
}
\end{figure}

We can understand the nature of these density structures from Fig.~\ref{Fig:PDtneq0_2}.
In the local density approximation, the spatial dependence comes from the local chemical potential.
The starting point of the trajectories of the chemical potentials ($\mu_{\alpha}^0$) for $^{87}$Rb varies with changes in the chemical potential for $^{41}$K.
In the case of Figs.~\ref{Fig:DP}(b) and (d), the trajectory of the chemical potential passes through the phase boundary of the $(n_1,0)$ state and the $(0,n_2)$ state, as represented by the dotted and dash-dotted lines in Fig.~\ref{Fig:PDtneq0_2}.
On the other hand, in the case of the nesting structure, the chemical potentials start from the $(n_1,0)$ region, pass through the $(0,n_2)$ region, and enter the $(n_1,0)$ region again, as shown by the dashed line in Fig.~\ref{Fig:PDtneq0_2}.
By the addition of more $^{41}$K atoms, the trajectory changes to the dash-dotted line.
The difference in the structure of the density distribution is caused by the difference in trajectories.

\section{Conclusions}
In this study, we investigated the properties of two-species Bose-Bose mixtures in an optical lattice using a mean-field approximation.
We derived the analogous expression for the phase boundary between the $(n_1,0)$ state and the $(0,n_2)$ state by comparing the ground energy in the $t_{\alpha}=0$ and $U_{12}\to\infty$ limit.
Using this phase boundary, we clarified the conditions for the spatial phase separation of Bose-Bose mixtures when both components are in the MI phase.
These results depend on the on-site interactions and occupation numbers.

We also studied the density distribution of $^{87}$Rb-$^{41}$K mixtures in a combined potential of a parabolic trap and an optical lattice using local density approximation.
We estimated on-site interactions and hopping amplitudes using experimental data \cite{Catani}, and calculated the densities and superfluid order parameters.
We fixed the number of $^{87}$Rb atoms and changed the number of $^{41}$K atoms. 
We found that phase separation occurs even for $^{87}$Rb-$^{41}$K mixtures with $^{87}$Rb as the SF phase and $^{41}$K as the MI phase.
We clarified that $^{87}$Rb is localized at the center of a parabolic trap and $^{41}$K encircles the $^{87}$Rb component.
For a special case, we found a nesting structure, as shown in Fig.~\ref{Fig:DP}(c).
These results are understood by considering the phase diagram in the $\mu_1/U_1$-$\mu_2/U_2$ plane of the Bose-Bose mixtures given in Fig.~\ref{Fig:PDtneq0}.
These results should be observed experimentally.

{\bf Acknowledgment}

We gratefully acknowledge valuable discussions with I. Danshita, S. Konabe and E. Arahata.
This research was supported by the Academic Frontier Project 2005 of MEXT.

\newpage %Just because of unusual number of tables stacked at end
%\bibliography{BEC}% Produces the bibliography via BibTeX.

\end{document}